\pdfoutput=1
\documentclass{sadhana}
\usepackage{graphicx,amsmath,bm}
\usepackage{color}
\usepackage{graphicx}

\begin{document}
\title{Galaxy Rotation Curve Measurements with Low Cost 21 cm Radio Telescope}
\author{Arul Pandian B\textsuperscript{1,4,*}, Ganesh L\textsuperscript{2}, Inbanathan S.S.R\textsuperscript{1,3}, Ragavendra K B\textsuperscript{4}, Somashekar R\textsuperscript{4} \and  Prabu T \textsuperscript{4}}
\affilOne{\textsuperscript{1} Post Graduate and Research Department of Physics, The American College, Madurai\\}
\affilTwo{\textsuperscript{2} Yadhava College, Madurai\\}
\affilThree{\textsuperscript{3} Department of Applied Science, The American College, Madurai\\}
\affilFour{\textsuperscript{4} Raman Research Institute, Bangalore\\}
\twocolumn[{
\maketitle
\begin{abstract}
    Probing the Universe with atomic hydrogen 21 cm emission is a fascinating and challenging work in astronomy.  Radio telescopes play a vital role in detecting and imaging these faint signals.  Powerful radio telescopes are complex to construct and operate. We have built a simple, low-cost 21 cm radio telescope primarily for educational training purposes.  The design uses a custom horn antenna, ready-to-use radio-frequency components, and a software-defined radio module.  The telescope operates efficiently from a rooftop in a city environment. Using this telescope, we have conducted observations and successfully detected the 21 cm line emissions from the different directions of our galactic plane.  Based on the Doppler-shift observed in these measurements, we have successfully derived the Galactic rotation velocity (rotation curve)  in those directions.   The paper presents the details of the telescope construction, 21 cm observation, and the Galactic rotation curve derivation.
\end{abstract}
\msinfo{16 April 2021}{23 August 2021}{03 January 2022}
 }]


\setcounter{page}{1}
\corres
\volnum{1}
\issuenum{1}
\monthyear{aa 2022}
\pgfirst{1111}
\pglast{1111}
\doinum{111.aaaa/1111}

\articleType{}
\markboth{author name for running head}{short title for running head}
\section{Introduction}
\begin{figure}[!ht]
        \centering
        \includegraphics[width=0.99\linewidth]{ 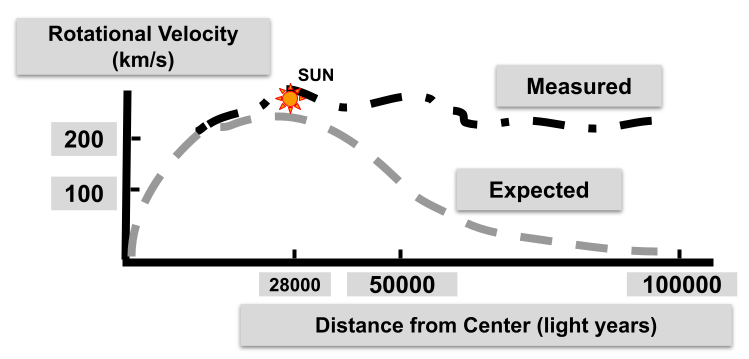}
             \caption{Cartoon depicts how the velocities at different radial distances differ from the theoretical expectations for our Galaxy. At high radial distances beyond about 28000 light-years from the galactic centre, velocities are higher than early theoretical predictions. Our solar system is at this distance of 28000 light-years (8.5 kpc) from the Galactic centre. Our observations presented here measure the rotational velocities beyond 8.5 kpc distance.}      \label{GeneralDiagram}
    \end{figure}

    The first-ever detection of neutral hydrogen ( {\emph {HI} }) from space created a new interest in radio telescope-based astronomy observations. H. C Van De Hulst (1942) theoretically predicted the hyperfine transition of  {\emph {HI} }. Subsequently, Harold  Ewen and Edward Purcell of Harvard University made the first experimental detection of the emission from the interstellar  {\emph {HI} } regions (1951). The  {\emph {HI} } emissions having an intrinsic wavelength of 21 cm at around 1420 MHz originate from distant places and directions in a galaxy, travel through the interstellar clouds and appear on ground-based radio telescope observations at Doppler-shifted frequencies \cite{kulkarni1988}.  An investigation into these frequency-shift renders details of the arrangement and relative velocities of the emission region.  The maps made of these  {\emph {HI} } emissions revealed the spiral structure of the Milky Way \cite{kraus1966}. The underlying mass distribution of the spiral structure influences the orbital velocity of the Galaxy at different radial positions.  Interestingly, the  {\emph {HI} } observations also revealed that at higher radial distances, the galactic arms are moving at a higher rotational velocity than expected for the known mass distribution in those positions, thus inferring the presence of unknown and otherwise undetected dark matter in the galaxies \cite{kellermann1988}. Figure \ref{GeneralDiagram} presents a cartoon to depict how the rotational velocities deviate from the expected rates for our Galaxy. 
    
    Today, there are many radio telescopes, such as the Giant Metrewave Radio Telescope near Pune in India, Parkes in Australia, Greenbank telescope in the USA, and the Jodrellbank telescope in the UK, are routinely performing 21 cm based observations. While these giant telescopes are very sensitive and capable of performing complex observational tasks, people have been building simple radio telescopes for educational experiment purposes. Building such telescopes provides training opportunities across various interdisciplinary areas such as electronics, antennas, signal processing, programming and astronomy. Efforts from Patel et al. and MIT Haystack observatory are prior such efforts \cite{low-cost_21cm} \cite{SRT}. In this work, we demonstrate a) constructing one such 21 cm radio telescope, b) how to observe, c) interpret the measurements and d) derive the rotation curve of our Galaxy.  

    We have built a simple, low-cost 21 cm radio telescope primarily for educational training purposes. We present a comprehensive account of the 21 cm radio telescope design details in the paper.  The design consists of a custom horn antenna, ready-to-use radio-frequency components, a software-defined radio receiver module, a computer for recording and analysing the data. With the telescope mounted on a rooftop, meridian transit observations of the galaxy were made. Transit positions were calculated using the standard calculator tools available in the public domain. A narrow radio-frequency (RF) band was sampled around 1420 MHz, and the data were collected in a frequency-switched mode at each position. Average spectrum corresponding to the sky positions were obtained and analysed to get the  {\emph {HI} } emission profiles. The emissions profiles are further processed to obtain the Galactic rotation velocity for the observed positions.  A velocity vector projection method is used to estimate the galactic rotation curve presented in the paper. Our measurements are also compared with the existing data \cite{mass} \cite{McGaugh}.  
    
\section{Paper Outline}
    Section 3 introduces the galaxy rotation curve calculation from the  {\emph {HI} } observations.  Section 4 provides the 21 cm receiver design details of a horn antenna, amplifiers, filters, data capture with software-defined radio, and the software for data acquisition and analysis.  Section 5 provides the observation and data analysis details. Section 6 presents our results with a discussion. Section 7 concludes by outlining the future scopes.
    
\section{Milky Way Rotation Curve} 
    Our Milky Way galaxy has spiral-shaped arms on a flattened disk with a bulging centre with a large number of stars. Milky Way has four major spiral arms, and our solar system is in one of the minor arms called Orion arm. The Solar System is at about 8.5 kpcs away from the centre of the galaxy.
    This arm rotates with an average velocity of about 220 km s\(^{-1}\). The galactic centre contains interstellar clouds that absorb visible light and hence are not optically visible. However, they are transparent to radio waves and hence the galaxy structure is known mostly from the radio studies of the neutral hydrogen.   
    
    Hydrogen is the most abundant element in the interstellar medium (ISM). From filling factor estimations, it accounts for about 75 percent of the Byronic mass in ISM. Within a radius of 10 kpcs from the Sun, our galaxy is estimated to contain about  \(4.8\times10^9\) M\(_\odot\) of  {\emph {HI} } \cite{kellermann1988}. The Neutral hydrogen ( {\emph {HI} })  atoms are spread all over the galactic disk and also seen in the low-density regions of the galaxy, Most of the neutral hydrogen remains on a flat disk. And hence it is one of the key tools to determine the shape of a Galaxy.
    
    \begin{figure}[H]
        \centering
        \includegraphics[height=6cm,  width=1\linewidth]{ 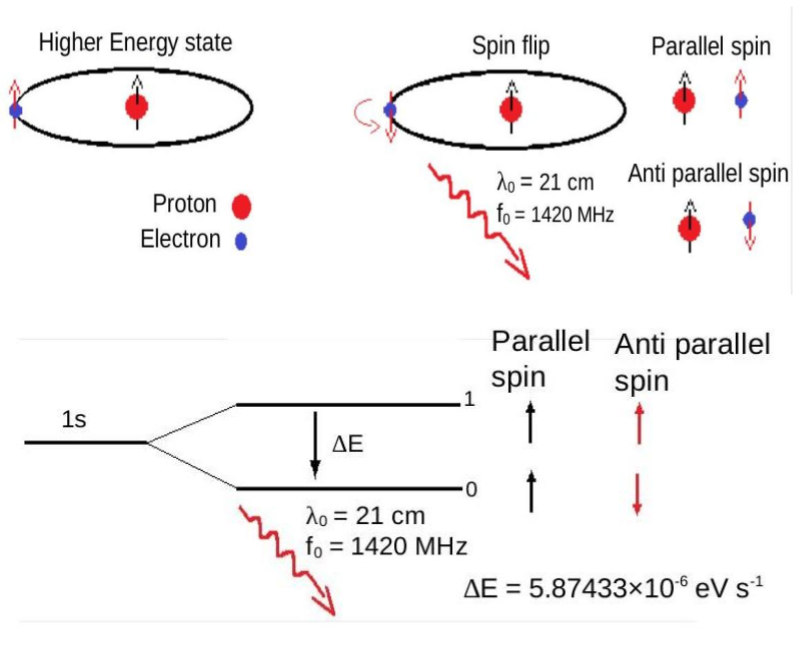}
        \caption{  Cartoon depicting how a spin-flip event generates the 21 cm emission from the hydrogen atom. During a spin-flip event,  an electron spontaneously changes its spin orientation back to the lower energy configuration by emitting a small energy equivalent to the energy difference between the two spin states.}
         \label{fig:GeneralDiagram1}
    \end{figure}

    The electron and the proton in  {\emph {HI} } atom form a tiny magnetic dipole. It will have a small amount of energy that varies according to its spin orientation. If their spin is parallel to each other (triplet state) then the energy will be higher. If their spin is anti-parallel (singlet state),energy will be lower. The energy difference between these two states \(\Delta\)E is about $5.88\times10{^{-6}} eV$
    \cite{marr2015} \cite{griffiths1982hyperfine}.
    The amount of energy emitted while the spin-flip transition is corresponding to the frequency   \(\upsilon\), 
       
    \begin{equation}
        \upsilon = \frac{\Delta E}{h} \approx1420 MHz.
    \end{equation}  where, \(h\) is Planck constant. The wavelength corresponding to this frequency is about 21 cm. This hydrogen line emission is popularly known as 21 cm line emission \cite{rohlfs2013}.  This process is illustrated in figure \ref{fig:GeneralDiagram1}.

    This spin-flip process is rare because, once a hydrogen atom got excited,  it would typically wait for about 11 million years before the next such spontaneous transition.   However,  we can observe the 21 cm line emissions in short-duration observations. This is mainly because a radio telescope beam, samples huge volumes of neutral hydrogen atoms from the galaxy. A significant number of transition happens in that observed volume of neutral hydrogen atoms.
     
    The  {\emph {HI} } frequency observed on an earth based observation is different from the rest frequency, because of the Doppler shift arising from the Galactic rotation. The relation between rest frequency \(f_0\)  and and the Doppler shifted frequency \(f\)  is given by    
    \begin{align}
        \label{Doppler shif formula}  
        \frac{f}{f_0} &= \Bigg[\frac{c + V_0}{c + Vr}\Bigg]   &             
            Vr &= \Bigg[\Bigg[{\frac{f}{f_0}}\Bigg][c + V_0]\Bigg] - c
    \end{align}
        where, \(c\) is the velocity of the light, 
        \(V_0\) is the velocity of the observer, 
        \(V_r\) is the source's velocity relative to the observer.  For \(V_0\), we can assume the sun's velocity around the milky-way. \(V_r\) is considered as negative for objects moving towards us (blue shift) and positive for objects moving away from us (red shift).
        Based on this directional Doppler shift  measurements of the  {\emph {HI} } emissions the  rotation curve for a Galaxy can be derived \cite{kraus1966} \cite{marr2015} \cite{zeilik1998introductory} . 
        
        Our Milky-Way Galaxy has a disk-like barred spiral shape. The spiral arms extend about 100,000 light-year distances from the Galactic centre. Our Solar system is located at a distance of about 28,000 light-years from the Galactic centre. Galactic arms have  {\emph {HI} } clouds that co move along with the spiral arms. A Galaxy rotation curve shown in figure \ref{GeneralDiagram} illustrates the variation in the orbital velocity of the galaxy at different radial distances from the galactic centre \cite{dickey1990hi}. The  {\emph {HI} } clouds move around the galaxy in a circular path, but each at different radial velocities. In radio telescope observations,  we will encounter different radial velocities at different distances at each pointing towards the Galactic arms. Hence, we will sample a wide range of line of sight velocities resulting in a complex but unique shape for the observed  {\emph {HI} } profiles \cite{kraus1966} \cite{kellermann1988}.   Figure \ref{fig:tangentialpoint} illustrates this situation, with an example  {\emph {HI} } profile with an imprint of different velocity information of the line-of-sight is shown in the figure insert. We can then use these  {\emph {HI} } profiles to derive the rotation curve of the Galaxy \cite{marr2015}.  
        
        The left side picture in figure \ref{fig:tangentialpoint} illustrates a top-level view of the Galaxy, showing emissions from four  {\emph {HI} } clouds at locations A, B, C, and D. These clouds are moving around the Galaxy, and a line-of-sight (thick line with an arrow-head) intercepts these locations. Each location A to D has a different radial velocity and an independent orbit with the Galactic centre. An observation made along this line of sight will result in sampling  {\emph {HI} } emissions emerging at different velocities.  The emissions arrive with varying Doppler shifts resulting in a  {\emph {HI} } profile with components corresponding to A, B,C and D as shown in the figure insert (right side).   
       
        Inside the Solar-system radius of our  Galaxy, the rotation curve can be derived using a tangent-point method. Outside the Solar-system radius, a velocity vector method can be used.  We have used the velocity-vector method in our measurements because we are mainly interested in  outside our solar system, but for completeness, both methods are presented here  \cite{mass} \cite{zeilik1998introductory} \cite{sofue}. 
        
    \subsection{Tangent-Point method}
       If the  {\emph {HI} } emission is measured inside the Solar-system radius the tangent-point method can be used to derive the rotation curves \cite{mass}.   Suppose the motion of an object is relative to the Sun. In that case, the movement towards or away from the Sun is called radial velocity \(V_r\), and the motion perpendicular to the direction of the Sun is called tangential velocity \(V_t\). The combination of the two motions is called the spatial velocity of the object. The radial component of the velocity is responsible for a Doppler shift of the spectral lines that can be determined directly, even if the distance is unknown.

    \begin{figure}[!ht]
        \centering
        \includegraphics[ width=8 cm ]{ 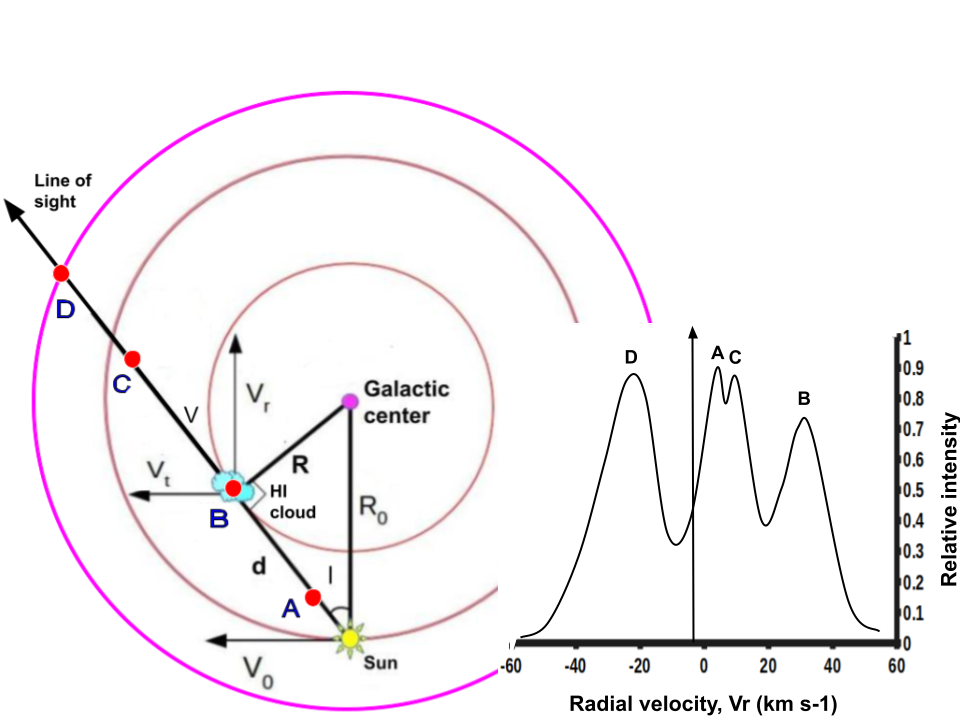}
        \caption{The Solar system is located at a distance of \(R_0\) from the Galactic centre. Inside the solar circle formed by \(R_0\), a line of sight in every direction will have a perpendicular line \(R\) at the tangent positions.  The position of the clouds is shown by the dots (A, B, C, D).  The  {\emph {HI} } profile observed along the line of sight is shown in the insert. It can be noted that the intensity of the profile is independent of R. Our experiments focus on measuring the rotation curve beyond the solar circle (i.e., beyond 8.5 kpc) and hence uses the velocity-vector method of estimation.}
        \label{fig:tangentialpoint}
    \end{figure} 
    
       The Sun is located from the galactocentric distance of \(R_0\) approximately 8 kpc in our galaxy.   The tangent-point method to obtain the rotation curve is useful for the radial distances \(R < R_0\).       Inside this solar circle, as illustrated in figure \ref{fig:tangentialpoint}, the galactic disk has tangential points at which the rotation velocity \(V\) is perpendicular to the distance of \(R\) from the galactic centre (GC). The line of sight velocity at distance R from the galactic centre can be calculated using the relation,
    \begin{equation}
        \label{tangent velocity} V(R) = V_r(max) + V_0 \sin{(l)}  
    \end{equation}
       where, \(V_r\) is the  {\emph {HI} } cloud's radial velocity,  \(V_r(max)\) is the maximum velocity at the tangent point, \(V\) is the cloud's circular orbital velocity. \(V_0\) is the circular orbital velocity of the sun and \(l\)  is the galactic longitude of  {\emph {HI} } cloud along the line of sight (LOS). The distance \(R\) to the tangent point is given by,
    \begin{equation}
         \label{Distance1} R = R_0 \sin{(l)} 
    \end{equation}
        where, \(R_0\) is the distance from the sun to the GC and d is the tangent point distance from the Sun \cite{zeilik1998introductory}.
        
    \subsection{Velocity-Vector method}  
\begin{figure}[!ht]
        \centering
        \includegraphics[ width=1\linewidth]{ 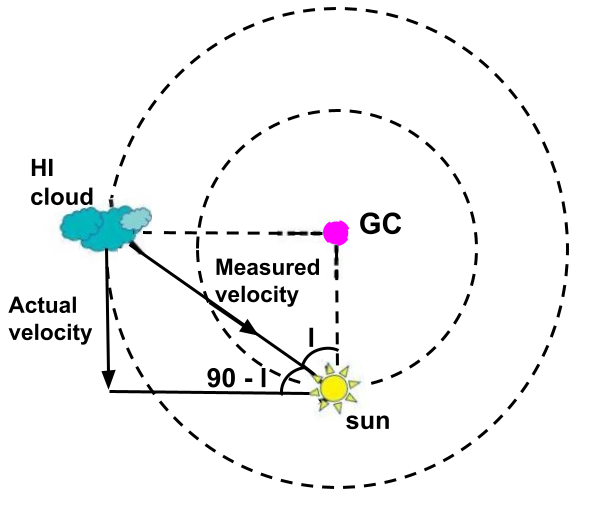}
        \caption{ Our work uses the velocity-vector method to estimate the rotation curve. The geometry concerning the relative arrangement of the Galaxy positions is shown in the figure. Velocity measured in the observation differs from the actual due to the projection effect. The actual  {\emph {HI} } cloud velocity  \(V_r\) of the cloud is calculated using the relation shown in equation (5).} 
        \label{velocity_vector}
    \end{figure}

        If the  {\emph {HI} } emission is measured outside the Solar-system radius the velocity vector method can be used to derive the rotation curves \cite{mass}. Using the geometry shown in figure \ref{velocity_vector} we can derive the relative radial velocity from the Doppler-shifted velocities measured on the Earth.
    
        The relative radial velocity of the emission can be derived using the relation,
    \begin{equation}
         V_r = \frac{Measured\ \ velocity}{\sin(90 - l)} 
    \end{equation}
        Relative radial velocity in-terms of angular velocity,
    \begin{equation}
        \begin{split}
            V_r = \Bigg[\frac{V}{R}-\frac{V_0}{R_0}\Bigg] \ \ R_0  \sin{(l)} \\
            V_r = (\omega - \omega_0)\ \ R_0 \sin{(l)}\\
            V_r = A\ \ d \sin{(2l)} \\
        \end{split}
    \end{equation}
        where, V is the velocity of the  {\emph {HI} } cloud, R is the distance from the Galactic centre, \(V_0\) is the circular orbital velocity of the Sun, \(R_0\) distance from the Galactic centre to the Sun, {\emph l} is the galactic longitude of  {\emph {HI} } cloud along the line
of sight (LOS) and d are the distance from Sun to the  {\emph {HI} } cloud (see figure \ref{fig:tangentialpoint}).
        
        \(\omega\) is angular velocity at tangential point, \(\omega_0\) is angular velocity of the Sun, derived for the Oort constant, A \cite{zeilik1998introductory}. 
        
    \begin{equation}
        \begin{split}
            \label{oort radial} 
            d = \frac{V_r}{A \sin{(2l)}}
        \end{split}
    \end{equation}
        
        Using \(\omega\) and d, we can calculate Relative Tangential Velocity,  
    \begin{equation}
         \begin{split}
             V_t = \Bigg[\frac{V}{R} - \frac{V_0}{R_0} \Bigg]\ \ R_0 \cos{(l)} \ \ - d\  \omega \\ 
             V_t = (\omega - \omega_0)\ \ R_0 \cos{(l)} - d\  \omega  
        \end{split}
     \end{equation}
     \begin{equation}
         \label{oort tangential} V_t = d \ \ (A \cos{2l}+B) 
     \end{equation}
        where, A and B are the Oort constants. \\
        Radial velocity is given by
    \begin{equation}
         \label{Rotational Radial velocity} U_r =V_r + V_0 \sin{(l)}
    \end{equation}
        Tangential velocity is given by
     \begin{equation}
         \label{Rotational Tangential velocity} U_t =V_t + V_0 \cos{(l)}
     \end{equation}
        Total velocity is given by
    \begin{equation}
        \label{Total velocity} V = \sqrt{U_r^2 + U_t^2} 
    \end{equation}
        The distance from galactic centre to  {\emph {HI} } cloud is given by
    \begin{equation}
        \label{Distance} R = \sqrt{R_0 ^2 + d^2 - 2R_0d \cos{(l)}}
    \end{equation}
        where, R is the distance from the galactic centre to the  {\emph {HI} } cloud on the line of sight, V is the velocity of the  {\emph {HI} } cloud at R. The values of R and V are used to draw the galaxy rotational curve that we have presented from our observations in Table.\ref{table1} and in figure 14.

\section{Front-End RF Receiver Design Details}
\begin{figure}[!ht]
        \centering
        \includegraphics[ width=1\linewidth]{ 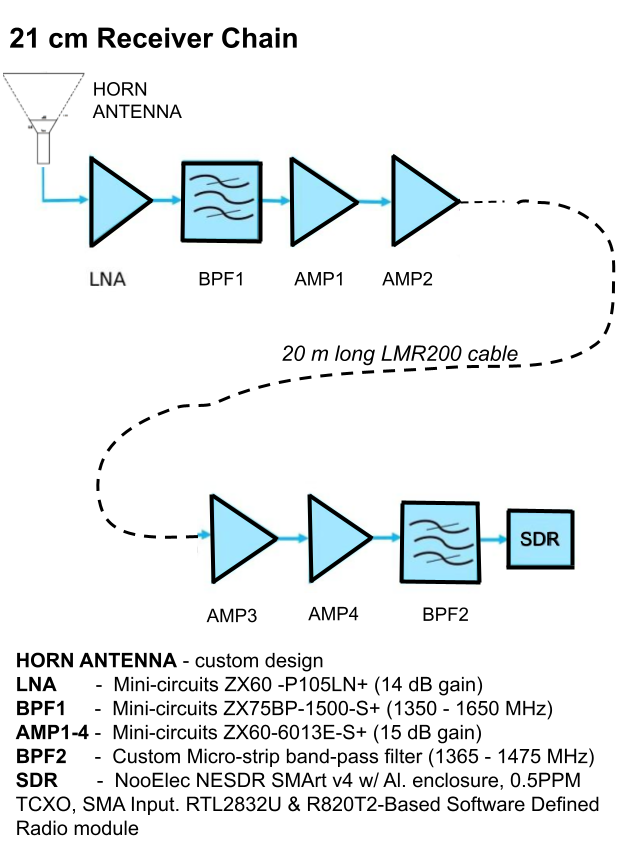}
        \caption{The main functional part of the 21 cm radio telescope is the front-end RF receiver chain. It consists of a horn antenna, one set of low-noise amplifiers (LNA), bandpass filters (BPF1 and BPF2), post amplifiers (AMP1 to AMP4). The horn antenna and one of the bandpass filters (BPF2) are custom designed for this work. The outputs from the RF receiver chain feeds to a software-defined radio (SDR) module.}
        \label{fig_rx_chain}
    \end{figure}

    In this section, we present the design details of the 21 cm telescope RF receiver system shown in figure \ref{fig_rx_chain} consists of a horn antenna, low-noise amplifier, bandpass filters and amplifiers. The horn antenna and one of the bandpass filters (BPF2) are custom designed for this work.
    
        The receiver-chain's figure of merit mentioned in the noise-figure is about 2.16 dB with the main contributing element to this value there being the LNA \cite{kraus1966}. Corresponding instantaneous noise floor estimated for the receiver is about -111 dBm. We need to apply sufficient averaging to achieve the higher sensitivity required for the observations. Details of the sensitivity calculation are presented in the appendix.
    
    \subsection{Horn Antenna}
        Horn-antenna couples the electromagnetic radio emission to the electrical circuit. We used a custom made single polarisation pyramidal horn antenna having 30 degrees of beam-width. The antennas feed length and the back-shot positions can be adjusted to operate it over a narrow range of frequencies. 
    
\begin{figure}[H]
        \centering
        \includegraphics[ width=1\linewidth]{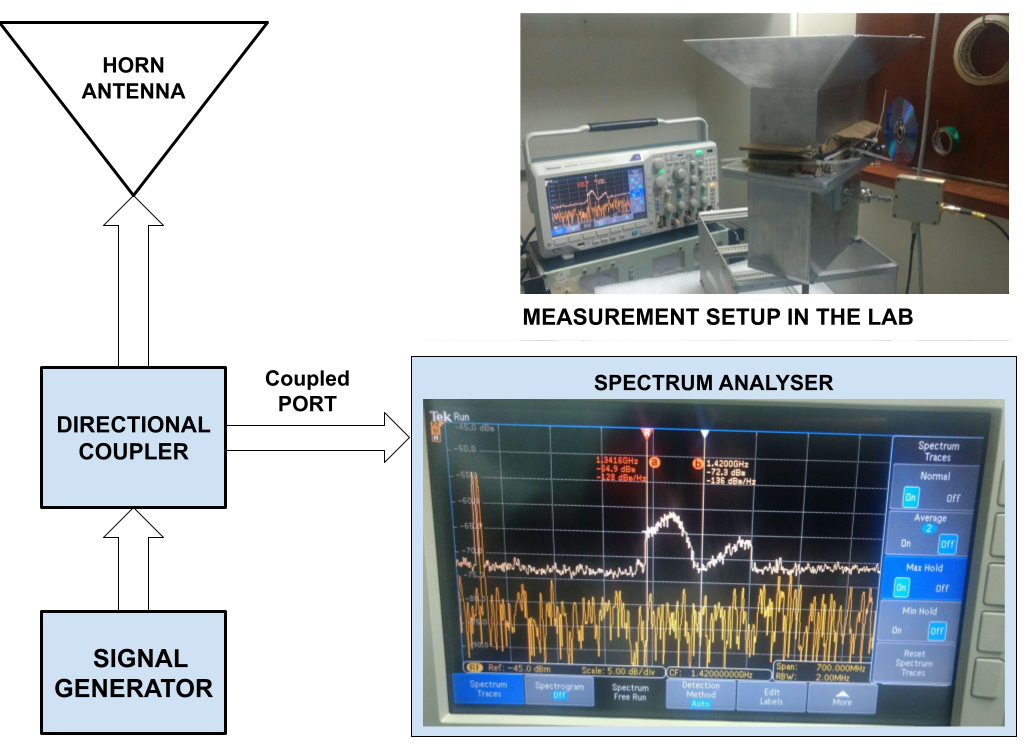} 
        \caption{A block diagram of the horn antenna characterization setup is shown on the left. The spectrum analyzer output after tuning the antenna at 1420MHz frequency is shown on the right. The figure on the top shows the laboratory setup depicting the arrangement of the horn antenna, directional coupler and spectrum analyzer.}
        \label{fig_tuning}
    \end{figure}

        The horn is tuned for an optimal performance at 1420 MHz using simple laboratory tools \footnote{Example: https://www.youtube.com/watch?v=vVuMYCdlsZw}. The arrangement consisted of a directional coupler and a frequency generator as shown in figure \ref{fig_tuning}. One port of the coupler was connected to the spectrometer to measure the reflected signal power from the Antenna. The other port was connected to the antenna. The third port of the coupler was fed with radio frequency tones around 1420 MHz. Frequency from a signal generator. Optimal response of the antenna is achieved when a characteristic dip in the spectrum appears as seen in the spectrum analyzer display shown in figure \ref{fig_tuning} lower right side, which corresponds to the sensitive reception band of the horn antenna. This tuning required adjusting the horn back-shot position and varying the feed probe length.
     
\begin{figure}[H]
        \centering
        \includegraphics[ width=1\linewidth]{ 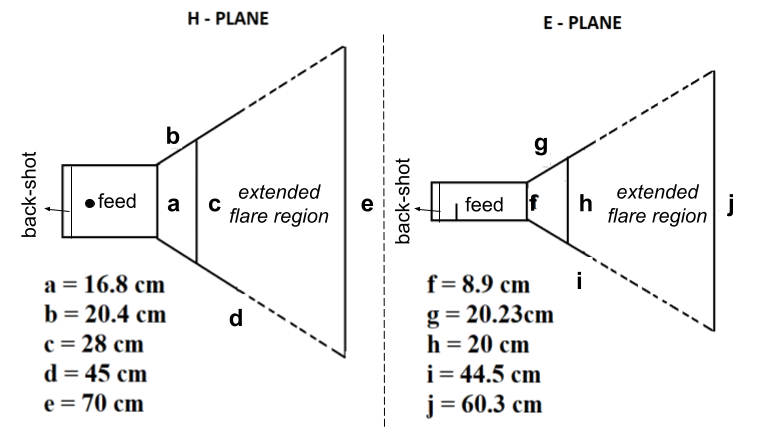}
        \caption{A block diagram of the horn antenna characterization setup is shown on the left. The spectrum analyzer output after tuning the antenna at 1420MHz frequency is shown on the right. The figure on the top shows the laboratory setup depicting the arrangement of the horn antenna, directional coupler and spectrum analyzer.}
        \label{antenna_dimentions}
    \end{figure}

       Since, we used an existing horn of smaller aperture available from the laboratory and extended its flare portion to suit the 21 cm observations. It was made of aluminum, having a back-shot, feed mount, and a flare with a dimension of ”a b c” and ”g f h” as shown in the figure\ref{antenna_dimentions}.   This flare portion was extended in the H-plane to the dimensions ”a-d-e” and in the E-plane to the extent  ”f-i-j” to achieve the desired higher gain.

    \begin{figure}[ht]{
        \centering
        \includegraphics[height= 3cm, width= 3cm ]{ 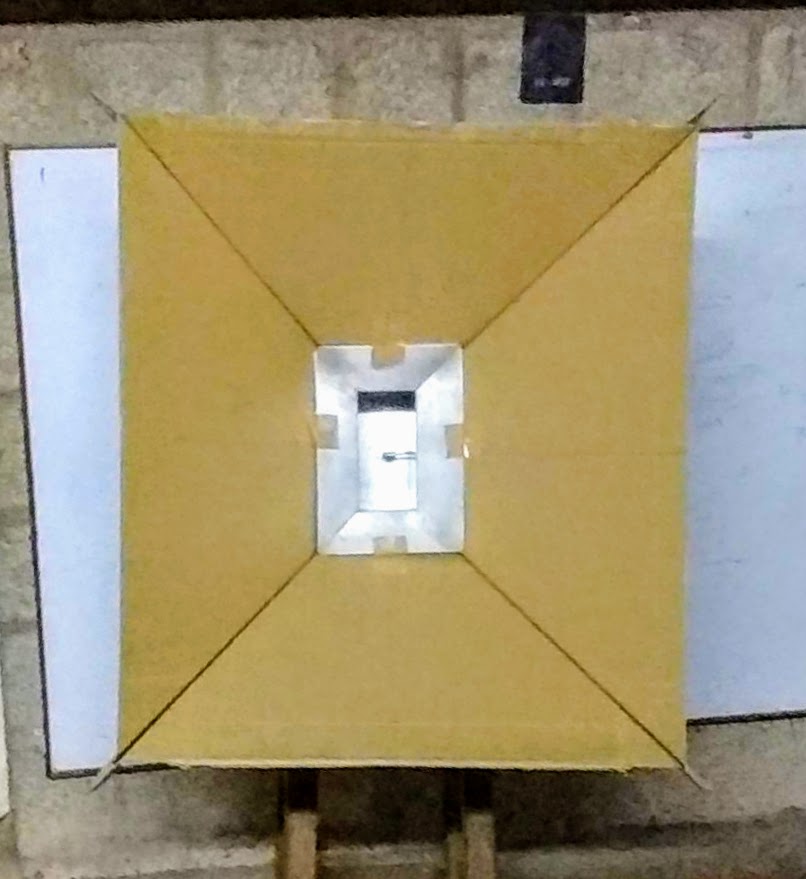}
        \hfill
        \includegraphics[height= 3cm, width= 5.2cm ]{ 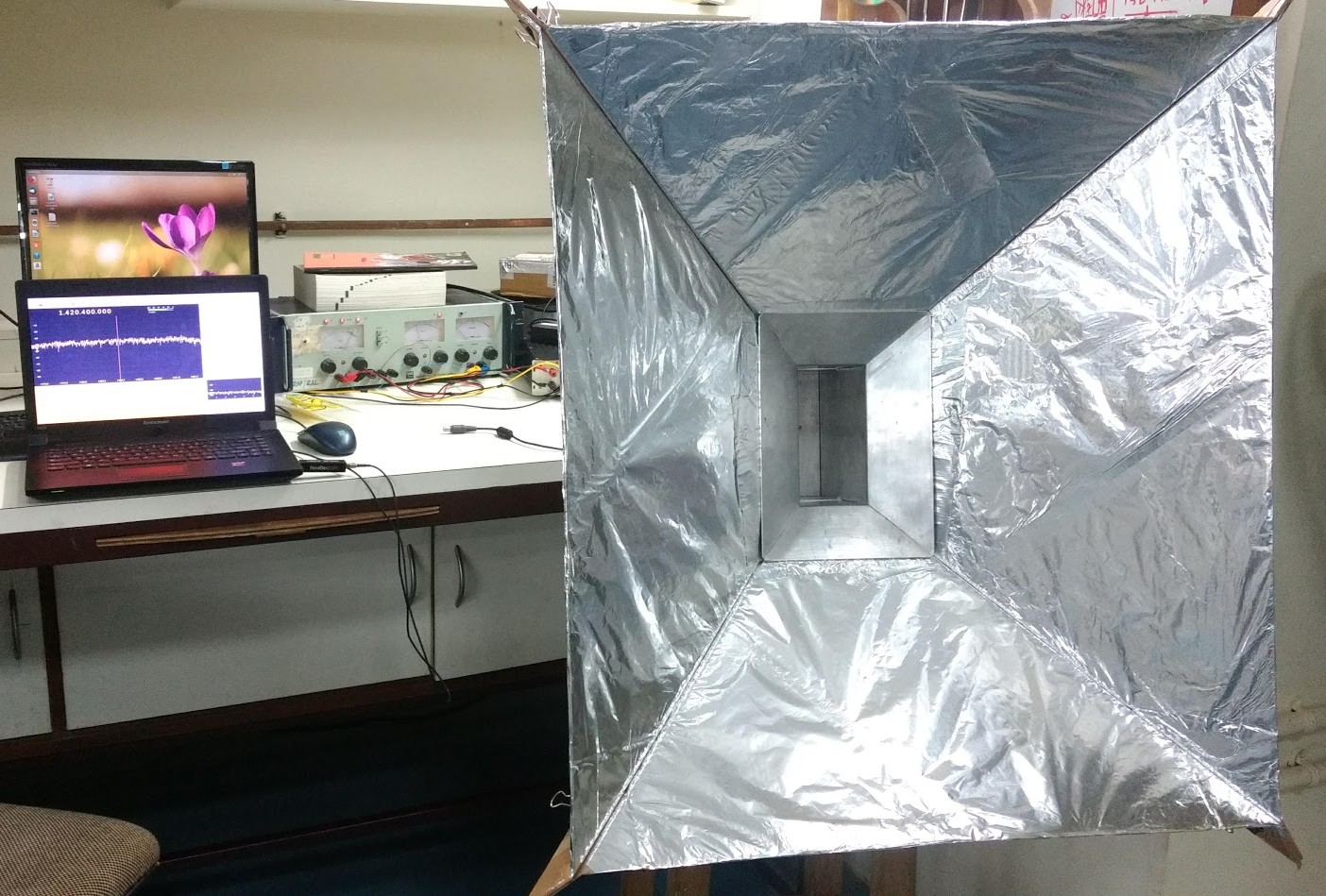}}
        \caption{Pictures show the horn antenna after card-board flare extension (left) and after application of Aluminium-foil reflecting surface (right).}
        \label{fig_flare}
    \end{figure} 
    
\begin{figure}[!ht]
        \centering
        \includegraphics[ width=1\linewidth]{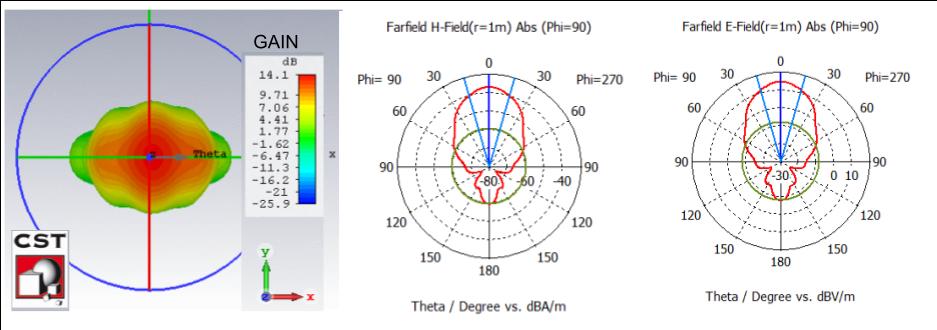}
        \caption{We have simulated horn antenna response using the CST (R) software. The beam patterns obtained for the flare extended horn antenna is shown in the figure. The pictures show antenna gain patterns along the top, H-plane and E-planes. The plots show that suitable high gain is achieved over a 30 degrees wide beam.}
        \label{fig_directivity}
    \end{figure}

    Dimension for the flare's extension was calculated by expanding angles of H-plane and E-plane about 72.5 for the H-plane and 71.0 for the E-plane. Thus the dimensions extended to give a flare exit-width of 70 cm in H-plane and 60 cm in E-plane. Cardboard was used as a base material in the extended region with an aluminum foil cover on the top for conduction. A 10-micron thick aluminum foil was used to provide sufficient skin-depth needed (about 2.3 microns) for the 1420 MHz signal as shown in figure \ref{fig_flare}.  We estimate the gain of the horn at 1420 MHz after the flare-extension as 13.3 dBi, which very closely matched the CST ® software simulation results presented in figure \ref{fig_directivity}.     Gain pattern,  H-field polar plot and E-field polar plot of the flare extended horn-antenna are shown in the left, middle and right figures, respectively. Simulated antenna responses are given in figure \ref{fig:s11plot}.  The antenna provides a collecting area of 0.42 \( m^{2}\) \cite{kraus1966} \cite{solid_angle}. 

    \subsection{RF Receiver chain Electronics } 
    
        The receiver electronics shown in figure \ref{fig_rx_chain} needs to be sensitive enough to detect the weak radio emission from the celestial sources that are typically much lower than \(10^{-20}\) watts.   
    
\begin{figure}[!ht]
        \centering
        \includegraphics[ width=1\linewidth]{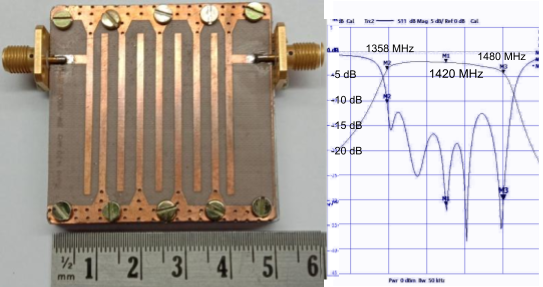}
        \caption{ The picture on the left shows the micro-strip custom filter developed for use in the RF receiver chain (BPF2 in figure \ref{fig_rx_chain}). The filter is etched on a double-sided PCB. It operates in the L-Band with a bandpass optimized with a -3 dB bandwidth of 110 MHz centred at 1420 MHz. It is a 9th order inter-digital Chebyshev type filter. The bandpass response plot shown on the left indicates the signal transmission loss and the input reflection coefficient. Details of this filter design and implementation considerations are presented in the appendix figure \ref{fig:BPFresponse}.}
        \label{fig:FilterPhoto}
    \end{figure}

        It consists of a low noise amplifier followed by bandpass filters and amplifiers. A low noise amplifier with a noise figure of 2 dB is used. The bandpass filters restrict out band radio frequency interference from contaminating the measurements. The first bandpass filter (BPF1) allows signals between 1350 MHz and 1650 MHz thus preventing GSM signals from contaminating the receiver. The two amplifiers (AMP1 and AMP2) provide an overall gain of about 30 dB and enable transmitting the signal through a long (about 20 m) co-axial cable to the laboratory. The signal transmission over the cable attenuates the signals by about 10 dB. The second part of the receiver located in the laboratory has two amplifier stages that further amplify the signal and compensate for the attenuation suffered by the signal during the cable transmission. The second bandpass filter BPF-2 used in the receiver is a microstrip-based filter. It operates over a 110 MHz band centred at 1420 MHz. The details of the constructions is provided in figure\ref{fig:FilterPhoto}. 
    
\begin{figure}[H]
        \centering
        \includegraphics[ width=1\linewidth]{ 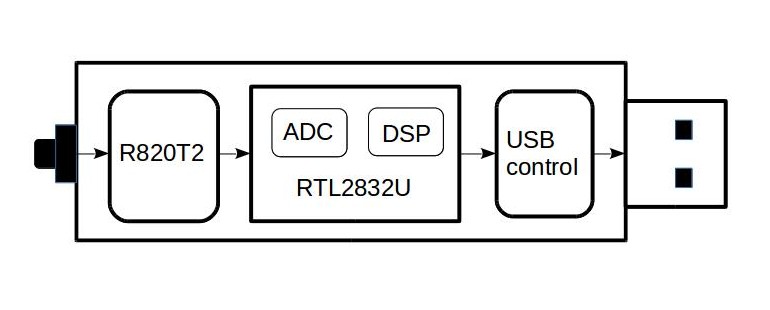}
        \caption{The functional blocks of the Software-Defined Radio (SDR) is shown in the figure. The main functional blocks are front end RF-amplifiers, filters, ADC and digital signal processing blocks with a USB based PC interface. The SDR forms the last module in our RF front-end receiver chain shown in figure\ref{fig_rx_chain}. We have used the NooLec RTL SDR (R) module for data digitization and recording purposes. We have tuned the SDR to digitize a narrow 1 MHz wide band around 1420 MHz. The frequency switched mode data collection was performed by shifting the SDR sampled band from the nominal centre frequency of 1420 MHz to a nearby centre frequency of 1420.7 MHz.}
    \label{fig:sdr_module}
    \end{figure}

        Then the bandpass processed signal is passed to a Software Defined Radio (SDR) module  (figure\ref{fig:sdr_module}) for digitisation and recording. Then the acquired data is processed using the analysis programs  developed specifically for this purpose \cite{github}. 
         

    \subsection{Data Acquisition}
        For the data acquisition and digitization, we used Software-Defined Radio (SDR). It allows us to specify the mode of operation, frequency band, sampling rate, and gain by our requirements.  It is a ready-to-use device, available on market at different costs. In this work, we choose NESDR SMArt v4 because its operating range is within our desired frequency.
        This commercial Software-Defined Radio module can be operated to digitize up to 2.4 MHz band over a range of centre frequencies between 25 MHz-1750 MHz. It has an 8-bit complex analog to digital converter and a processor. We defined a 1 MHz band with a centre frequency of 1420 MHz for the data acquisition. 
        We have used a frequency-switched method to collect the data during the observations \cite{dickey1990hi}. The sky positions were observed at two closely spaced centre frequencies. The SDR centre frequency (\(f_2\)) was switched between 1420.0 MHz (\(f_1\)) and 1420.7 MHz (\(f_2\)) respectively. Frequency switched data set was useful to remove the pass-band ripples introduced by the SDR internal filters. 
        The SDR analog input gain was maintained at 25 dB. A sampling rate (\(f_s\)) of 1 MS/s was used in the SDR. The data were collected for 10 s at each of the two frequency settings for every sky positions as show in figure \ref{fig_skyposition}. 
    
\section{Observation Plan and Data Analysis}

\subsection{Observation}

\begin{figure}[H]
        \centering
        \includegraphics[ width=1\linewidth]{ 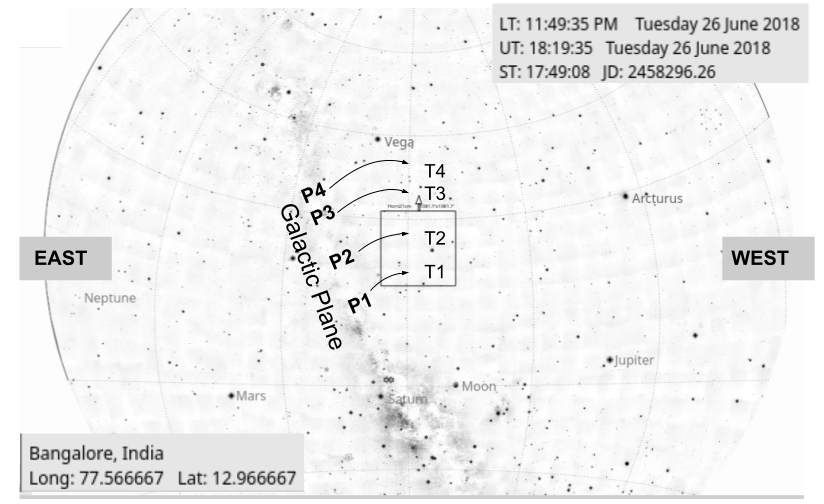}
        \caption{Observed sky positions are indicated as P1 to P4 in the sky-map shown in the figure. We have placed the antenna towards the zenith to acquire data when the sky positions P1 to P4 crosses the antenna beam at times T1 to T4.}
        \label{fig_skyposition}
    \end{figure}

       We have planned to observe the 21 cm emissions between the Galactic longitude l=30 and l=90 degrees \cite{marr2015}. We assume our antenna is pointing at galactic latitude b= 0 on the galactic plane.  Corresponding positions in the sky are between P1 to P4 in figure\ref{fig_skyposition}.  The signals from this region of the Galactic arm would be detected with a positive Doppler shift due to their velocity to-wards us.  Observing the 21 cm emission from this region (i.e., beyond 8.5 kpc or beyond 28000 light-years from the Galactic centre) is also of interest  as we can see the galactic rotation curve deviating from the {\emph Keplerian} rotation curve induced by the observed mass distribution in our Galaxy \cite{Roy2019Discrepancy}.   

        We have fixed our horn antenna to observe the positions P1 to P4 during their meridian transit. These sky positions transits at different times T1 to T4. Whenever a given position (P1 to P4) is at zenith, we made frequency-switched measurements for 10 seconds. A software tool (kstar\footnote{kstars: https://edu.kde.org }) was used to obtain the meridian transit time on the observation day at the telescope site ({\em Bengalore}) latitude  \cite{github}.
        
        The nominal galactic latitudes observed by this arrangement are  50$^{\circ}$, 59$^{\circ}$, 70$^{\circ}$, 75$^{\circ}$. At each of these positions, the horn antenna receives signals from over a 30$^{\circ}$ wide region, corresponding to  +/-15$^{\circ}$ from the nominal galactic latitudes, thus covering the 35$^{\circ}$ to 90$^{\circ}$ latitude range.

    \begin{figure}[!ht]
        \centering
        \includegraphics[ width=1\linewidth]{ 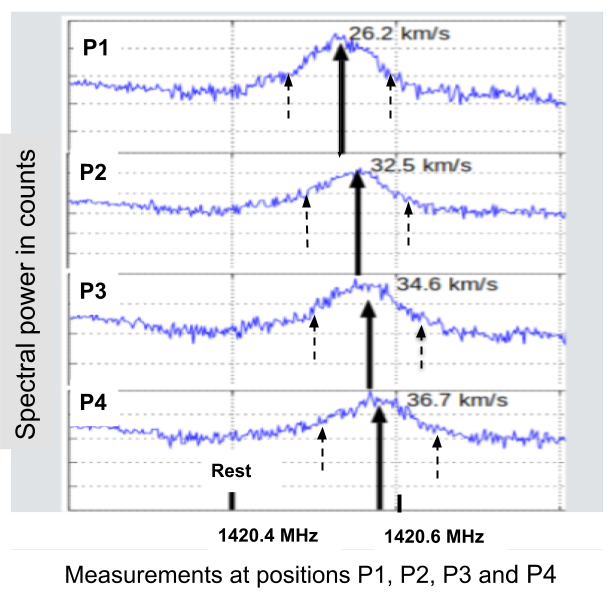}
        \caption{  {\emph {HI} } emission profiles obtained for the four sky positions P1 to P4 (figure \ref{fig_skyposition}) are shown here. Observed  {\emph {HI} } emissions has contributions from the nominal galactic latitude l\(\pm\)15$^{\circ}$. Hence profiles spread over a narrow band due to the contributions from a wide region with different velocities. The velocity calculations need to account for these extended contributions. We show the boundaries of the spectral region in the profiles using vertical arrows. The two short arrows indicate the left and right side frequencies, while the thick arrow in the middle shows the median frequency of the spectral profile considered in our calculations. We consider only these three discrete frequencies at each profile to simplify the calculations.}
        \label{spec}
    \end{figure}

    \begin{figure*}[!ht]
        \centering
        \includegraphics[height= 11cm,width= 16cm]{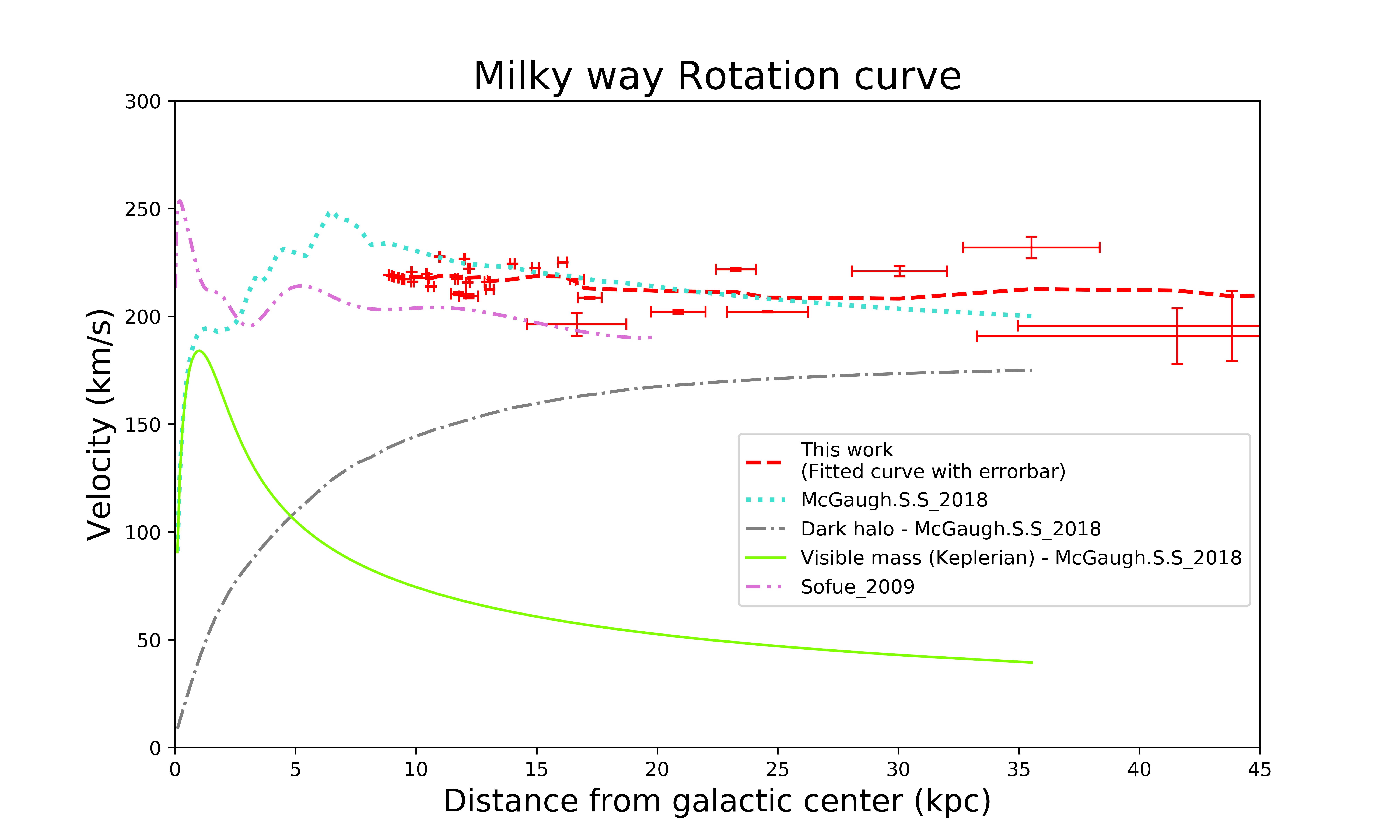}
        \caption{Rotation curve measurements from our observation using the 21 cm Radio telescope is presented here with error bars. The red-dashed line is a fit to our measurements.  The Keplerian relation for the known mass is shown by the "Visible mass (Keplerian) -McGaugh.S.S\_2018" curve (green color).  It corresponds to the velocities expected for a non-rigid body that obeys the equation (\ref{Rotation_curve}). For comparison, we have overlaid existing rotation curve measurements from McGaugh.S.S\_2018  (Blue- Dotted line) and Sofue\_2009 (Pink- double dotted dashed line) in the figure.}
        \label{result_plot}
    \end{figure*}

     
        The horn antenna was mounted on rooftop of the laboratory in the Bangalore city environment and manually pointed to the different sky positions during their meridian transit times. The SDR was operated in the frequency switched mode to collect 1 MSPS data for 10 s at each position, for each of the two frequency settings: (\(f_1 \& f_2\)) = 1420.0 MHz  \& 1420.7 MHz,respectively \cite{marr2015}. The SDR records eight-bit complex samples in an unsigned binary format. 
        
    \subsection{Data analysis}
    
        The data was processed using the gnu octave 512-point FFT to produce 1.953 kHz resolution average spectra for each 10 s data \cite{github}. Thus we obtained two average spectra for each sky position, one for each frequencies.  The frequency switched spectra are subtracted from each other to produce the 21 cm spectrum shown in figure \ref{spec} . And it is used to calculate the rotational velocity from Doppler-shifted frequencies . Since our telescope covers a large area in the sky because of large beamwidth, we took three frequency components from each spectral plot.  Three arrows indicate frequency components taken from two extreme ends with the centre frequency of the signal in figure \ref{spec}. We also took two extreme-end position measurements with nominal galactic latitude for sky pointing positions. Hence, we have generated nine measurements at each position and tabulated them in columns 2 to 4 in Table \ref{table1}.

        Using the Doppler-shifted frequency and galactic longitude, we calculated relative radial and tangential velocities with equations (6) and (9). The radial velocities \(U_r\) of the  {\emph {HI} } cloud shown in column 6 of Table. \ref{table1} are calculated using equation (10) and tangential velocity \(U_t\) in column 7 using the equation (11). Both are the velocity vectors of moving  {\emph {HI} } cloud in a circular orbit. The relative radial velocity\(V_r\) and galactic longitude \(l\) are used in equation (7) to derive the distance \(d\) from Sun to the object with  {\emph {HI} }. The net velocity \(V\) of the  {\emph {HI} } cloud is shown along column 9 is calculated by radial and tangential velocity using equation (12). The distance \(R\) of the the neutral hydrogen object from the galactic centre is calculated using equation (13). As per the IAU recommendations, the values of \(R_0\) and \(V_0\) used in the analysis are 8.5 kpc and 220km/s.  The values of \(R\) and \(V\) estimated from our measurements are given along columns 8 and 9 in Table \ref{table1}.
        
    \subsection{Error Analysis}
        The results can be affected by the following uncertainties:\\
    \begin{itemize}
        \item  For the data acquisition, We are using a commercial SDR made by NooElec that has an error in the crystal oscillator of 0.5 ppm. So the error could be approximately 7KHz when the crystal operates at 1420MHz.
        \item  Our antenna has a beamwidth of \(30^\circ \pm 2^\circ \). This pointing inaccuracy arises mainly from the beam asymmetry. This will also result in an error in the assumed galactic longitude(l).
        \item  For these observations, we assumed the galactic latitude (\emph{b}) be zero. Since our observation at each point lasts for 10's of seconds, the sky drifts, and hence the nominal latitude \emph{b}=0 assumptions will not hold for the full observing duration. This can add a small error to the calculations.
        \item  We manually pointed our telescope to the sky. Asymmetry in the positioning can  also make an error in the pointing.
      \item  The SDR may have a small frequency offset while sampling the signal. This offset will translate the profile frequencies with corresponding offset errors.
    \end{itemize}
    
         Among these various uncertainties, the dominant errors are 0.5 ppm error from the crystal oscillator and \(\pm 2^\circ\) pointing error from the antenna beamwidth asymmetry.  Hence we took these two errors in calculations. Results shown in Table \ref{table1} for the radial distance and velocity columns (8 and 9) include these error considerations.  
            
    \begin{table*}[!hpt]
        \centering
        \caption{Distance from galactic centre to  {\emph {HI} } cloud \(R\) and total velocity \(V\) computed using Velocity-Vector method }
        \setlength{\tabcolsep}{7pt} 
        \renewcommand{\arraystretch}{1.3} 
        \begin{tabular}{c c c c c c c c c } 
        \hline
        1 &2 &3 &4 &5 &6 &7 &8 &9 \\ [0.5ex]
        S.No &l(\(\theta)\) &b(\(\theta\))\(^{\star}\) &Frequency(MHz) &\(V_r\)(km/s) &\(U_r\)(km/s) &\(U_t\)(km/s) &R(kpc) &V(km/s) \\ [0.5ex] 
        \hline
        1 &35	&0	&1420.44	&-7.2	&171.3	&136.7    &8.8 $\pm$0.003  	&219.2 $\pm$0.037 \\
        2 &35	&0	&1420.52	&-23.9	&150.9	&161.0    &9.8 $\pm$0.008  	&220.7 $\pm$0.014 \\
        3 &35	&0	&1420.60	&-42.5	&128.3	&188.0    &10.9 $\pm$0.050    &227.6 $\pm$0.082 \\
        4 &50	&0	&1420.44	&-7.2	&130.1	&175.4    &9.0 $\pm$0.009  	&218.4 $\pm$0.077 \\
        5 &50	&0	&1420.52	&-23.9	&104.1	&191.3    &10.5 $\pm$0.026 	&217.8 $\pm$0.001 \\ 
        6 &50	&0	&1420.60	&-42.5	&75.2   &209.0    &12.1 $\pm$0.162 	&222.2 $\pm$0.221 \\ 
        7 &65	&0	&1420.44	&-7.2	&75.8	&202.4    &9.8 $\pm$0.0184  	&216.1 $\pm$0.184 \\
        8 &65	&0	&1420.52	&-23.9	&36.2	&209.3    &13.0 $\pm$0.045 	&212.4 $\pm$0.039 \\
        9 &65	&0	&1420.60	&-42.5	&-7.5 	&217.1    &16.6 $\pm$0.286 	&217.2 $\pm$0.041 \\
        10 &44	&0	&1420.44	&-7.4 	&147.8	&161.2    &8.9 $\pm$0.005     &218.7 $\pm$0.023 \\ 
        11 &44	&0	&1420.53	&-28.0	&119.2	&184.5    &10.4 $\pm$0.022    &219.7 $\pm$0.017 \\ 
        12 &44	&0	&1420.63	&-48.6	&90.5 	&207.8    &12.0 $\pm$0.304 	&226.7 $\pm$0.741 \\
        13 &59	&0	&1420.44	&-7.4 	&98.8 	&193.6    &9.4 $\pm$0.018     &217.3 $\pm$0.034 \\
        14 &59	&0	&1420.53	&-28.0	&58.7	&207.6    &12.1 $\pm$0.081 	&215.7 $\pm$0.073 \\
        15 &59	&0	&1420.63	&-48.6	&18.7 	&221.5    &14.9 $\pm$1.135 	&222.3 $\pm$0.882 \\ 
        16 &74	&0	&1420.44	&-7.4 	&33.5 	&207.8    &11.7 $\pm$0.032    &210.5 $\pm$0.103 \\ 
        17 &74	&0	&1420.53	&-28.0	&-41.2	&197.9    &20.8 $\pm$0.140    &202.2 $\pm$0.040 \\
        18 &74	&0	&1420.63    &-48.6	&-115.9	&188.0    &30.0 $\pm$1.965 	&220.9 $\pm$2.339 \\
        19 &55	&0	&1420.44	&-7.4	&113.1	&186.2    &9.2 $\pm$0.014 	&217.9 $\pm$0.001 \\
        20 &55	&0	&1420.54	&-30.1	&73.6 	&204.5    &11.6 $\pm$0.124 	&217.4 $\pm$0.270 \\ 
        21 &55	&0	&1420.64	&-50.7	&37.7 	&221.2    &13.9 $\pm$2.062 	&224.4 $\pm$5.278 \\ 
        22 &70	&0	&1420.44	&-7.4 	&53.4 	&207.1    &10.6 $\pm$0.054 	&213.9 $\pm$0.026 \\
        23 &70	&0	&1420.54	&-30.1	&-12.8	&208.3    &17.1 $\pm$0.496 	&208.7 $\pm$0.446 \\
        24 &70	&0	&1420.64	&-50.7	&-73.1	&209.4    &23.2 $\pm$8.307 	&221.8 $\pm$12.905 \\
        25 &80	&0	&1420.44	&-7.4 	&-4.8 	&196.2    &16.6 $\pm$0.090	&196.3 $\pm$0.058 \\ 
        26 &80	&0	&1420.54	&-30.1	&-135.3	&134.4    &41.5 $\pm$0.833    &190.8 $\pm$0.616 \\ 
        27 &80	&0	&1420.64	&-50.7	&-254.0	&78.3     &64.2 $\pm$13.985	&265.8 $\pm$42.636 \\
        28 &60	&0	&1420.44	&-7.4 	&95.0 	&195.3    &9.4 $\pm$0.025  	&217.2 $\pm$0.023 \\
        29 &60	&0	&1420.55	&-32.2	&45.5 	&211.1    &12.9 $\pm$0.394 	&216.0 $\pm$0.980 \\
        30 &60	&0	&1420.66	&-53.9	&2.1  	&225.1    &16.0 $\pm$2.062    &225.1 $\pm$5.278 \\ 
        31 &75	&0	&1420.44	&-7.4 	&28.0 	&207.4    &12.1 $\pm$0.105 	&209.3 $\pm$0.077 \\ 
        32 &75	&0	&1420.55	&-32.2	&-67.4	&190.5    &24.5 $\pm$1.689 	&202.1 $\pm$0.292 \\
        33 &75	&0	&1420.66	&-53.9	&-151.4	&175.7    &35.5 $\pm$8.875 	&231.9 $\pm$16.306 \\
        34 &80	&0	&1420.44	&-7.4 	&-4.8	&196.2    &16.6 $\pm$0.176	&196.3 $\pm$0.101 \\
        35 &80	&0	&1420.55	&-32.2	&-147.2	&128.8    &43.8 $\pm$2.827 	&195.6 $\pm$5.062 \\ 
        36 &80	&0	&1420.66	&-53.9	&-272.4	&69.6     &67.7 $\pm$14.865   &281.1 $\pm$46.388 \\ 
        \hline
        \end{tabular}
        \begin{tabular}{l}
            (1) Serial number,  (2) Galactic longitude \(l\),  (3) Galactic latitude \(b\),  (4) Doppler shifted frequency \(f\) at 10 KHz resolution, \\  
            (5) Relative velocity \(V_r\),  (6) Radial velocity \(U_r\),  (7) Tangential velocity (\(U_t\)): displayed up to single decimal place\\
            (8) Distance from galactic centre from  {\emph {HI} } cloud \(R\) and  (9) Total velocity \(V\).
            \(^{\star}\)observed position's nominal latitude 
        \end{tabular} 
        \label{table1}
    \end{table*}
    
\section{Results}

    Galactic rotational velocities at different galactic radius estimated using our 21 cm measurements are presented in the plot shown in figure \ref{result_plot}. The Keplerian rotation curve is expected for the visible mass and the rotation curve of the milky way measured by McGaugh.  S. S. and  Sofue.  Y. ,are presented as an overlay for a comparison \cite{McGaugh} \cite{sofue} \cite{sofue1}.The rotational velocity \(V\) from Table \ref{table1} columns 9 are plotted against the radial distance \(R\) column 8 in red dashed line with error bars in figure \ref{result_plot}. 
    
    When we equate force from Newton’s law of gravitation with centripetal force of the object with mass m \cite{sofue1}, 
        
        \begin{align}
            \label{Rotation_curve}  
            \frac{mV^2}{R} &= \frac{GMm}{R^2}    &     
                V &=\sqrt {\frac{GM}{R}} 
        \end{align}
        
    Where \(G\) is the gravitational constant. \(R\) is the radial distance. \(V\) is the rotational velocity. The function \(V(R)\) is the rotational velocity concerning distance from the galactic centre. The rotational velocity \(V\) is expected to be directly proportional to the visible mass distribution \(M\) (\(V \propto\ M\)) and inversely proportional to the radial distance \(R\) (\(V \propto\ R^{-{1/2}}\)) as the disk in differential rotation with decrease in \(V\) with increasing \(R\) \cite{mass}.
    
        \begin{align}
            \label{Rotation_curve1}  
            V &= \frac{2\pi R}{P}   &
            P &= \frac{2\pi R^{3/2}}{\sqrt{GM}}            
        \end{align}
    Where, \(P\) is the period of the rotation.
    From above relations, We can derive Newton's form of Keplerian third law,
        \begin{align}
            \label{Rotation_curve2}  
            P &= \frac{2\pi R}{V}            
        \end{align}
    
    From the kepler's third law, the orbital velocity  decreases over the radius increase. The Keplerian relation is shown by the "Visible mass (Keplerian) - McGaugh.S.S\_2018" curve (green color) in Figure \ref{result_plot}. It corresponds to the velocities expected for a non-rigid body that obeys the equation (\ref{Rotation_curve}). Our rotation curve measurements (red dash-line) are plotted along with error bars in the same figure. For comparison, we have overlaid exiting measurements from McGaugh.S.S\_2018  (Blue- Dotted line) and Sofue\_2009 (Pink- Double dotted dash line) in the figure. 
        
    As it can be seen that the existing measurements of rotational velocity curves differ from that of the {\emph  Keplarian} rotation curve shown, specifically beyond a radial distance of about 8.5 kpc. Rotational velocities observed beyond this distance are nearly very similar, resulting in a flat rotational velocity curve extending out to large radii of the milky-way. Based on the known physical laws, it can be inferred that the observed higher velocity is coming from the presence of an additional unknown mass, the dark halo.  It can be seen that in figure \ref{result_plot}, the Dark halo (Black- Dashed dot line) is significantly contributing to making the observed velocities nearly constant up to large radial distances of our Galaxy. This contribution is increasing towards the increase of galactic distance that can relate with mass in equation (14).
    Thus, the 21 cm based measurements provide a unique handle to sense the effect of otherwise undetected Dark matter in the galaxies.

\section{Future scopes}
    The 21 cm radio telescope receiver that we designed, worked well and we were able to repeat the observations a few times. We are now considering a few improvements for the receiver that can be accommodated in the future. Presently, we have operated the horn antenna in a simple non-rotating mount. We had to physically move the horn antenna, using external angle markings to point it to different declination directions for our meridian transit observations. The horn mount can be improved by incorporating one-axis (elevation) rotation with fixed-angle holds for easy positioning during repeated observations.  The flare extension that we have made uses card-board and aluminum foils. During rain, the card-board based flares get wet and spoiled. We have used plastic covers to protect them. The flare sheet can be replaced in future by rigid aluminum sheets. In the present flare extension, the flares are detachable as four sheets, and we found it was very convenient for transporting the horn antenna. If the flares were to be made in aluminum, it would be helpful to retain the detachable flare arrangement.   Our horn has a wide flare-angle resulting in side-lobe reception, which is not desirable for precision measurements. Future design will be useful to reduce the flare-angle, requiring further horn modelling effort using suitable antenna simulation software. The noise calculation presented in the appendix is based on theoretical values. We would like to do the actual measurements of the noise floor of the receiver using standard laboratory calibration tools. Could use calibrated receivers for any advanced experiments and for the flux estimation of the observed 21 cm emissions. Our present horn uses a single polarisation feed. We can incorporate a dual-polarized feed horn and perform more sensitive 21 cm measurements. It will also be possible to construct a feed horn array to perform phased array and interferometric mode observations of the 21 cm emissions. If we use multiple SDR for the phased array or the interferometric mode, then a suitable synchronisation method is to be evolved. Thus,  this work opens up new student-level observational experiments:  rotation curve measurement at different angular positions, improving the precision of measurements by using more giant horn antennas, longer integration time, improved spectral resolution, and gain calibration schemes. Also, an investigation into polarised emission study and the estimates of the mass distribution of the Galaxy are possible \cite{mass}. Thus, the receiver design presented here extends into additional experimental projects of interest and use to the science and engineering students.

\section{Conclusion}
     We have developed a simple radio telescope receiver to observe the 21 cm emissions.  The design of the receiver is mostly based on low-cost commercial ready-to-use components. We have used two  custom components: horn antenna and a micro-strip bandpass filter. The design details of the custom components are presented in detail.  We have completed the design and operated the telescope to detect the Doppler-shifted 21 cm line emission from specific directions of the galactic plane. Subsequently, we have calculated the Milky Way galaxy rotation curve from our measurements. We have also presented the detailed steps to calculate the rotational velocity from the 21 cm measurements. The computational steps are presented in detail, and step-wise calculations of the results are presented in a tabular form. The rotation curve derived from the measurement is presented as a plot overlaid with two other existing measurements.  We have also presented a comprehensive list of future scopes in work. The design and cost details of the telescope and the codes used in for data collection and analysis are archived in github \cite{github}. We see that the 21 cm radio telescope receiver that we have designed can be easily constructed and operated for 21 cm observations from a rooftop and in a city environment, making it a convenient radio telescope for introductory astronomy and engineering experiments by students.

 \section*{Acknowledgement}
        This work was supported by the Raman Research institute EEG department and the American College Madurai. We thank Nimesh Patel for the very useful discussions at the early stages of this work. We thank Raghunathan for the antenna-related discussions. Maghendran for his help with the coordinate conversion tool. We thank our colleagues from the EEG department for their valuable comments that greatly improved our work. We also thank the anonymous referees for their comments.

\newpage
\appendix
 
\section{Appendix}
    
    \begin{enumerate}
        \item Noise figure\\
            We used the Friis formula to calculate the total noise figure $NF_{total}$ of the Receiver chain \cite{kraus1966}. Every active element in the receiver chain will add noise to the signal. 
        \begin{equation}
            \label{noise figure} NF_{total} = NF_1 + \frac{NF_2 -1}{G_1}
            + \frac{NF_3 -1}{G_1 G_2} + ..... + \frac{NF_n -1}{G_1 G_2 ... G_n}      
        \end{equation}
            Here,\\
            NF is the noise figure of the individual elements. \\
            G is their gain.\\
            In this formula the first element decides the noise figure's maximum value. Therefore we used a low noise amplifier as the first element at the receiver chain. From this formula, we calculated the total noise figure of the receiver chain as 2.1675 dB.
        \item Minimum detectable signal\\
            The minimum detectable signal (MDS) is the minimum power level that can process by a receiver. It is also known as the noise floor of the system. It can also be defined as the input signal power required to give a particular SNR at the output.
        \begin{equation}
            \label{noise floor} MDS = 10log_{10} \Big(\frac{kT}{1mW}\Big) +  NF_{total} + 10log_{10}(BW)
        \end{equation}
            Where, BW is the band width of the receiver. We find that system noise floor for our receiver is -111 dBm.
    \end{enumerate}

    \begin{figure}[!ht]
        \centering
        \includegraphics[ width=1\linewidth]{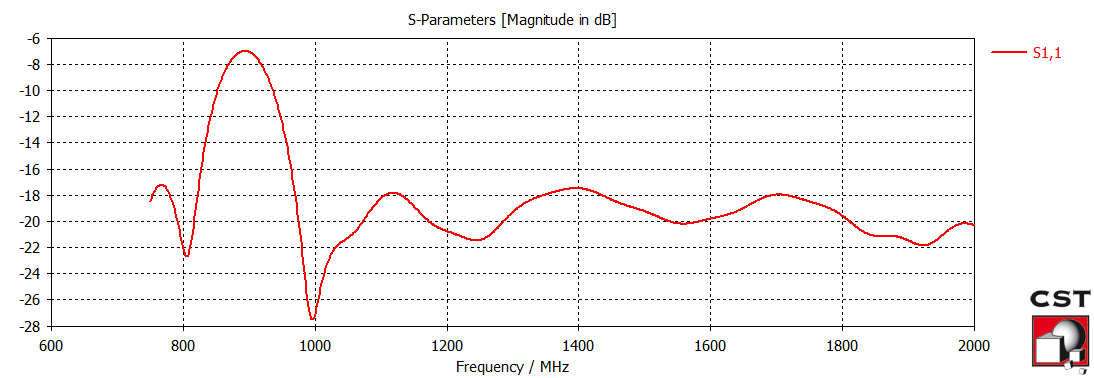}
        \caption{\textbf{Horn antenna return loss (S11) performance.} The CST software based simulation of the flare extended horn-antenna shows that the antenna can perform well beyond about 1000 MHz. It can also be noted that the antenna would perform poor below 1000 MHz. This poor performance at lower frequencies is desirable, as it will help to block some of the unwanted frequencies, especially the GSM mobile phone signals (around 900 MHz) from saturating the 21 cm receiver. GSM signals are typically very strong in a city environment and would contaminate the sensitive radio telescope when operated nearby. It can also be noted that S11 parameter value around 1420 MHz is better than -13 dB implying that a good performance from the antenna is expected for the 21 cm signal reception.}  
    \label{fig:s11plot} 
    \end{figure}

    \begin{figure}[!htp]
        \centering
        \includegraphics[ width=1\linewidth]{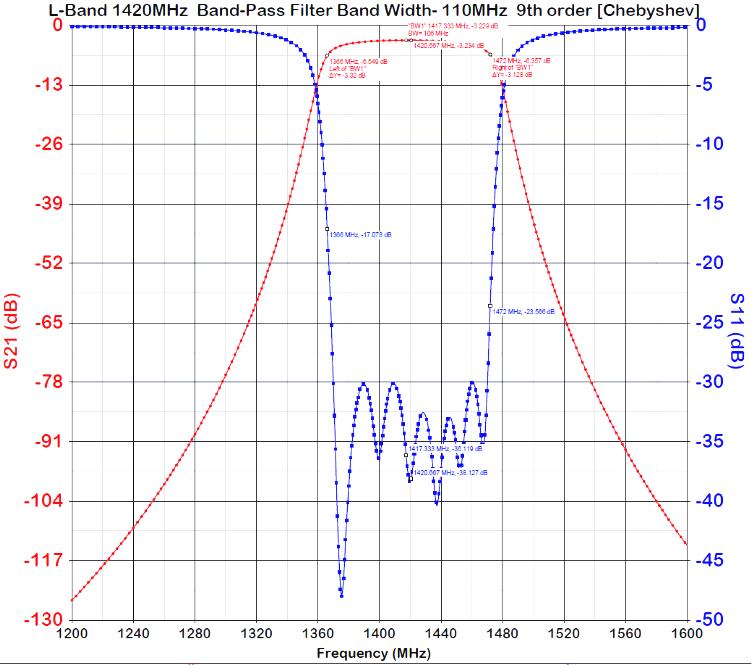}
        \caption{\textbf{Construction details and simulated response of the 1420 MHz bandpass of the Micro-strip filter}. This filter is a 9th order inter-digital Chebyshev micro-strip bandpass filter, which is implemented on 0.8mm di-electric thickness high frequency printed circuit board which is popularly known as ULTRALAM-2000, having di-electric constant \(\epsilon_r\) of 2.5 and loss tangent of 0.0022. The design of this filter is done in Keysight Genesys 10 CAD software. The inter-digital filter is a compact configuration consists of an array of nine TEM-mode transmission line resonators, each of which has an electrical length of 90\(^\circ\) at the mid-band frequency and is short-circuited at one end and open-circuited at the other end with alternative orientation. In general, the physical dimensions of the line elements or the resonators as indicated by the widths W1-W9. Coupling is achieved by way of the fields fringing between adjacent resonators separated by specified spacing. The grounding of micro-strip resonators, which is accomplished via holes. However, because the resonators are quarter-wavelength long using the grounding, the second pass-band of the filter is centred at about three times the mid-band frequency of the desired first pass-band, and there is no possibility of any spurious response in between. The measured frequency response of the implemented filter is shown in the left side. The design criteria for pass band and stop band attenuation are at -2 dB and 30 dB respectively, with the pass-band ripple of 0.01 dB. The optimized  -3 dB bandwidth of 110MHz centred at 1420 MHz.  It can be observed that the centre frequency is slightly deviated from 1420 MHz by 1 MHz on the higher side and the -3 dB band width is almost 120 MHz with a downwards slope of 2 dB. The rejection of -30 dB attenuation at 1340 and 152 0MHz frequencies with the bandwidth of 180 MHz has been achieved, which is the indicative of obtained form factor of the filter is in the order of 1.5.}
        \label{fig:BPFresponse}
    \end{figure}

\newpage

\end{document}